\documentstyle[multicol,aps,psfig,epsfig]{revtex}
\begin{document} 
\draft
\title{Stability of the doped antiferromagnetic state of the $t-t'$-Hubbard model }
\author{Avinash Singh\cite{avinash} and Haranath Ghosh\cite{ghosh} }
\address{Department of Physics, 
Indian Institute of Technology Kanpur - 208016, India}
\maketitle
\begin{abstract} 
The next-nearest-neighbour hopping term $t'$ is shown to stabilize 
the AF state of the doped Hubbard model with respect to 
transverse perturbations in the order parameter 
by strongly suppressing the intraband particle-hole processes.
For a fixed sign of $t'$, this stabilization is found to be 
significantly different for electron and hole doping,
which qualitatively explains the observed difference 
in the degree of robustness of the AF state 
in the electron-doped ($\rm Nd_{2-x}Ce_x Cu O_4$)
and hole-doped ($\rm La_{2-x}Sr_x Cu O_4$) cuprates.
The $t'-U$ phase diagram is obtained for both signs of the $t'$ term,
showing the different regions of stability and instability of the doped antiferromagnet. 
Doping is shown to suppress the $t'$-induced frustration due to the 
competing interaction $J'$. 
A study of transverse spin fluctuations in the metallic AF state 
reveals that the decay of magnons into particle-hole excitations yields 
an interesting low-energy result $\Gamma \sim \omega$ for magnon damping.
\end{abstract}
\begin{multicols}{2}\narrowtext
\section{Introduction}
The problem of determining the self-consistent 
mean-field (MF) state of an interacting electron system is 
generally equivalent to minimizing 
the ground-state energy function $E_G \{\Delta \}$ 
with respect to an assumed order-parameter set $\{\Delta\}$.
However, it may happen 
that when this set is expanded 
to include an additional component, 
the original self-consistent fixed point 
turns out to be a local maximum
with respect to this new direction in order-parameter space. 
In other words, 
the original self-consistent fixed point actually represents,
in such a case, 
a saddle point in the expanded order-parameter space. 

This is precisely the fate of the antiferromagnetic (AF)
state of the doped Hubbard model with
nearest-neighbour (NN) hopping $t$.
The self-consistent AF state is described by the gap parameter $\Delta_z$, 
related to the sublattice magnetization
$m_z$ in the $z$ direction by $2\Delta_z = m_zU$. 
This state was found to be unstable 
with respect to a new direction represented by 
the transverse component $\Delta_\perp $.
Inclusion of the transverse magnetization $m_\perp$ amounts to
spin twisting away from the  perfect AF alignment,
and the above instability is, in fact, towards the so-called 
spiral state,\cite{spiral1,spiral2,spiral3,spiral4}
characterized by an ordering wave vector ${\bf Q}$,
different from the AF wave vector $(\pi,\pi)$.
The spiral state is stabilized by an energy gain 
(for the doped system) resulting from an activation 
of the O($t$) hopping due to the 
slight twisting of neighbouring spins 
relative to the AF alignment. 
In terms of the deviation $\tilde{\bf q}  = {\bf Q} - (\pi,\pi)$
from the AF wave vector, the twist-induced hopping energy gain 
$\sim t \tilde{q} $ was found to compete with the 
twist-induced exchange energy loss $\sim J \tilde{q} ^2$,
resulting in an optimum spiral pitch for a given 
doping concentration. 
Thus, minimization of $E_G \{\Delta \}$, 
after expanding the order-parameter set 
to include the transverse component $\Delta_\perp$
(or equivalently the spiral wave vector ${\bf Q}$), 
results in a fundamentally new self-consistent fixed point.

In this paper we show that for a {\em positive} next-nearest-neighbour (NNN)
hopping term $t'$ in the Hubbard model (see Eq. 1 for our convention), 
the above instability is strongly suppressed for {\em hole doping},
and the self-consistent AF state of the doped $t-t'$-Hubbard model 
becomes stable for a range of doping concentration.\cite{mf}
However, for negative $t'$ and hole doping, we find that while
the AF state survives longitudinal perturbations in the order parameter,
it remains unstable with respect to transverse perturbations for any doping. 
To the extent the $t-t'$-Hubbard model is equivalent
to the three-band Cu-O Hamiltonian,\cite{tb1,tb2}
this result is particularly relevant to doped cuprates 
in which AF ordering is much more robust with respect to electron doping.
The electron doped cuprate $\rm Nd_{2-x}Ce_x Cu O_4$
retains AF order up to a doping concentration of 
about  12\%,\cite{electron}
whereas only 2\% hole concentration destroys AF order 
in $\rm La_{2-x}Sr_x Cu O_4$.
Since a particle-hole transformation maps the $t'$ model and hole (electron)
doping on the $-t'$ model and electron (hole) doping,
the positive (negative) $t'$ model and hole (electron) doping is appropriate to study
for the electron-doped compound $\rm Nd_{2-x}Ce_x Cu O_4$,
for which a negative $t'$ is usually assumed.

More significantly, a stable AF state 
of the doped $t-t'$-Hubbard model
provides a microscopic realization of the {\em metallic AF state}, 
in which the Fermi energy lies within a quasiparticle band. 
Indeed, metallic antiferromagnetism has been reported in 
$\rm \kappa - (BEDT-TTF)_2 X$,\cite{metaf1}
$\rm V_{2-x} O_3$,\cite{metaf2}  and $\rm NiS_{2-x} Se_x$.\cite{metaf3}
This allows us to quantitatively study  
spin excitations in the doped AF. Specifically, 
we evaluate the magnon damping $\Gamma$ arising from magnon decay into 
intraband particle-hole excitations across the Fermi energy.
Interestingly, we find that in the low-energy limit, 
the magnon damping $\Gamma \sim \omega$. 
This is of relevance 
in the context of phenomenological theories 
such as the nearly AF Fermi-liquid theory,\cite{af_flt}
put forward to explain the Knight shift experiments and 
describe the non-Korringa temperature dependence 
of the nuclear spin-lattice relaxation rate in doped cuprates. 

The need for more realistic microscopic models which include NNN hopping etc., 
has been acknowledged recently from 
band structure studies, photoemission data
and neutron-scattering measurements of high-T$_{\rm c}$
and related materials.\cite{nnn1,nnn2,nnn3,nnn4}
Estimates for $|t'/t|$ range from 0.15 to 0.5.
Effect of hole and electron doping on the commensurate spin ordering
have been studied for the $t-t'$-Hubbard model and applied to 
$\rm La_{2-x}Sr_x Cu O_4$ and $\rm Nd_{2-x}Ce_x Cu O_4$.\cite{chub}
Spin correlation function, incommensurability,
and local magnetic moments in the doped $t-t'$-Hubbard model
have been studied using the Quantum Monte Carlo method.\cite{duffy95}
At half filling, existence of a metallic AF phase has been suggested 
in $d=2$.\cite{duffy97}
The suppression of the perfect-nesting instability 
by the NNN hopping, and the critical
interaction $U_c$ vs. $t'$ phase diagram has been studied  
in $d=2,3$.\cite{hofstetter}
Magnon softening due to $t'$ and a significant enhancement in the  
low-energy spectral function due to single-particle excitations has
been observed.\cite{spectral}
Evolution of the magnon spectrum with $t'$ was studied, and the 
quantum spin-fluctuation correction to sublattice magnetization
in $d=2$ and the N\'{e}el temperature in $d=3$ were also evaluated.\cite{spectral} 
Finite-$U$ effects on competing interactions and frustration
were recently examined, and the magnetic phase diagram was  obtained
at half filling.\cite{phase}
Recently effects of nnn hopping have also been studied in 
one-dimensional systems involving chains and ladders.\cite{ghosh+santos}

\section{Metallic AF state}
We consider the $t-t'$-Hubbard model on a square lattice,
with NN and NNN hopping terms $t$ and $t'$
connecting sites $i$ to $i+\delta$ and $i+\kappa$, respectively:
\begin{equation}
H = 
-t \sum_{i,\delta,\sigma} ^{\rm NN}
a_{i, \sigma}^{\dagger} a_{i+\delta, \sigma}
-t' \sum_{i,\kappa,\sigma} ^{\rm NNN} 
a_{i, \sigma}^{\dagger} a_{i+\kappa, \sigma}
+  U\sum_{i} n_{i \uparrow} n_{i \downarrow} \; .
\end{equation}
In the following we set $t=1$.
As the $t$ and $t'$ terms connect sites of opposite and same sublattice,
respectively, the corresponding free-fermion energies
$\epsilon_{\bf k} = -2t(\cos k_x + \cos k_y)$ and
$\epsilon'_{\bf k}= -4t'\cos k_x \cos k_y$
appear in off-diagonal and diagonal matrix elements
of the HF Hamiltonian,
and in the two-sublattice basis\cite{st} we have
\begin{equation}
H_{\rm HF}^\sigma ({\bf k})=
\left [ \begin{array}{cc}
-\sigma \Delta + \epsilon'_{\bf k} & \epsilon_{\bf k}  \\
\epsilon_{\bf k} & \sigma \Delta+ \epsilon'_{\bf k}   \end{array} \right ]
= \epsilon'_{\bf k} \; {\bf 1} + 
\left [ \begin{array}{cc}
-\sigma \Delta & \epsilon_{\bf k}  \\
\epsilon_{\bf k} & \sigma \Delta  \end{array} \right ]
\end{equation}
for spin $\sigma$.
Here $2\Delta=mU$, where $m$ is the sublattice magnetization.
Since the $\epsilon' _{\bf k}$ term appears as a unit matrix,
the eigenvectors of the HF Hamiltonian remain unchanged
from the NN hopping case.
The fermionic quasiparticle amplitudes on the two sublattices A and B are 
given by the eigenvector $(a_{{\bf k}\sigma} \;\; b_{{\bf k}\sigma})$ of
the HF Hamiltonian matrix, 
and for spin $\sigma=\uparrow,\downarrow$
and the two quasiparticle bands $\ominus,\oplus$, we have\cite{st}

\begin{eqnarray}
& &(a_{{\bf k}\uparrow}^\ominus)^2 = 
(b_{{\bf k}\downarrow}^\ominus)^2 =
(a_{{\bf k}\downarrow}^\oplus)^2 = 
(b_{{\bf k}\uparrow}^\oplus)^2 = 
\frac{1}{2}\left ( 1 + \frac{\Delta}
{\sqrt{\Delta^2 + \epsilon_{\bf k} ^2} } \right ) \nonumber \\
& &(a_{{\bf k}\uparrow}^\oplus)^2 = 
(b_{{\bf k}\downarrow}^\oplus)^2 =
(a_{{\bf k}\downarrow}^\ominus)^2 = 
(b_{{\bf k}\uparrow}^\ominus)^2 = 
\frac{1}{2}\left ( 1 - \frac{\Delta}
{\sqrt{\Delta^2 + \epsilon_{\bf k} ^2} } \right ) . \nonumber \\
& &   
\end{eqnarray}

\begin{figure}
\vspace*{-70mm}
\hspace*{-38mm}
\psfig{file=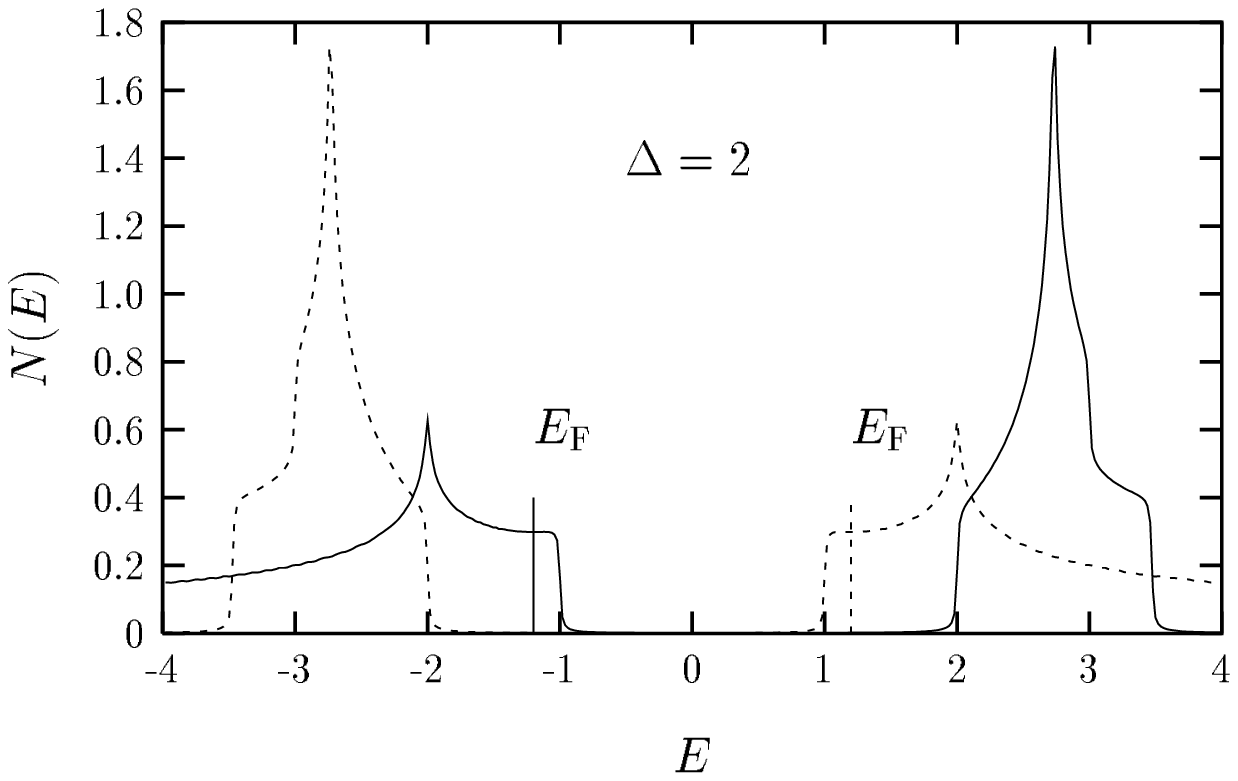,width=135mm,angle=0}
\vspace{-70mm}
\caption{
The AF-state electronic density of states (DOS) for positive $t'$
(solid) and negative $t'$ (dashed), 
showing the flat DOS near the Fermi energy for hole and
electron doping, respectively.}
\end{figure}
\noindent
In the strong coupling limit ($U>>t$),
the majority and minority densities
given above reduce to $1 - \epsilon_{\bf k}^2 /4\Delta^2 $
and $\epsilon_{\bf k}^2 /4\Delta^2 $, respectively. 

The quasiparticle energies corresponding to the eigenvalues of the
HF Hamiltonian are, however, modified to
\begin{equation}
E_{{\bf k}\sigma}^{\pm}=\epsilon'_{\bf k} 
\pm \sqrt{\Delta^2 + \epsilon_{\bf k} ^2} \; ,
\end{equation}
the two signs $\pm$ referring to the lower and upper quasiparticle bands.
The band gap is thus affected  by the NNN hopping term,
and it progressively decreases 
as $2\Delta - 4t'$ in the weak coupling limit.
The corresponding density of states
is shown in Fig. 1 for both positive and negative $t'$,
showing the drastic reduction near the upper and lower
band edges of the lower and upper bands, respectively.

We first consider the case of positive $t'$ and hole doping;
the same results hold for negative $t'$ and electron doping.
The highest electronic energy levels of the lower AF band 
$E_{\bf k}^{\ominus}=\epsilon'_{\bf k} 
- \sqrt{\Delta^2 + \epsilon_{\bf k} ^2}$
correspond to $(k_x,k_y) = (\pm \pi,0)$ and  $(0,\pm \pi)$,
and with doping hole pockets develop around these points.
In this region of ${\bf k}$ space 
constant-energy surfaces are nearly (semi) circular, 
as shown in Fig. 2.
Therefore, for low hole doping 
the Fermi surface consists of these four semi circles. 
The radius $a$ of the {\em hole} Fermi circle 
is related to the doping concentration $x$.
The ratio of the total area inside the Fermi circles $(2\pi a^2)$
to the total Brillouin zone area $(4\pi^2)$ yields the fraction
of unoccupied states $(x)$, so that $a^2 = 2\pi x$.

We now examine the AF states which lie near the
Fermi energy $E_{\rm F}$, and will therefore contribute to the
particle-hole processes at low energies.
For ${\bf k}$ states lying close to a Fermi circle,
say near the point $(-\pi,0)$, 
$k_x, k_y$ can be

\begin{figure}
\vspace*{-60mm}
\hspace*{-28mm}
\psfig{file=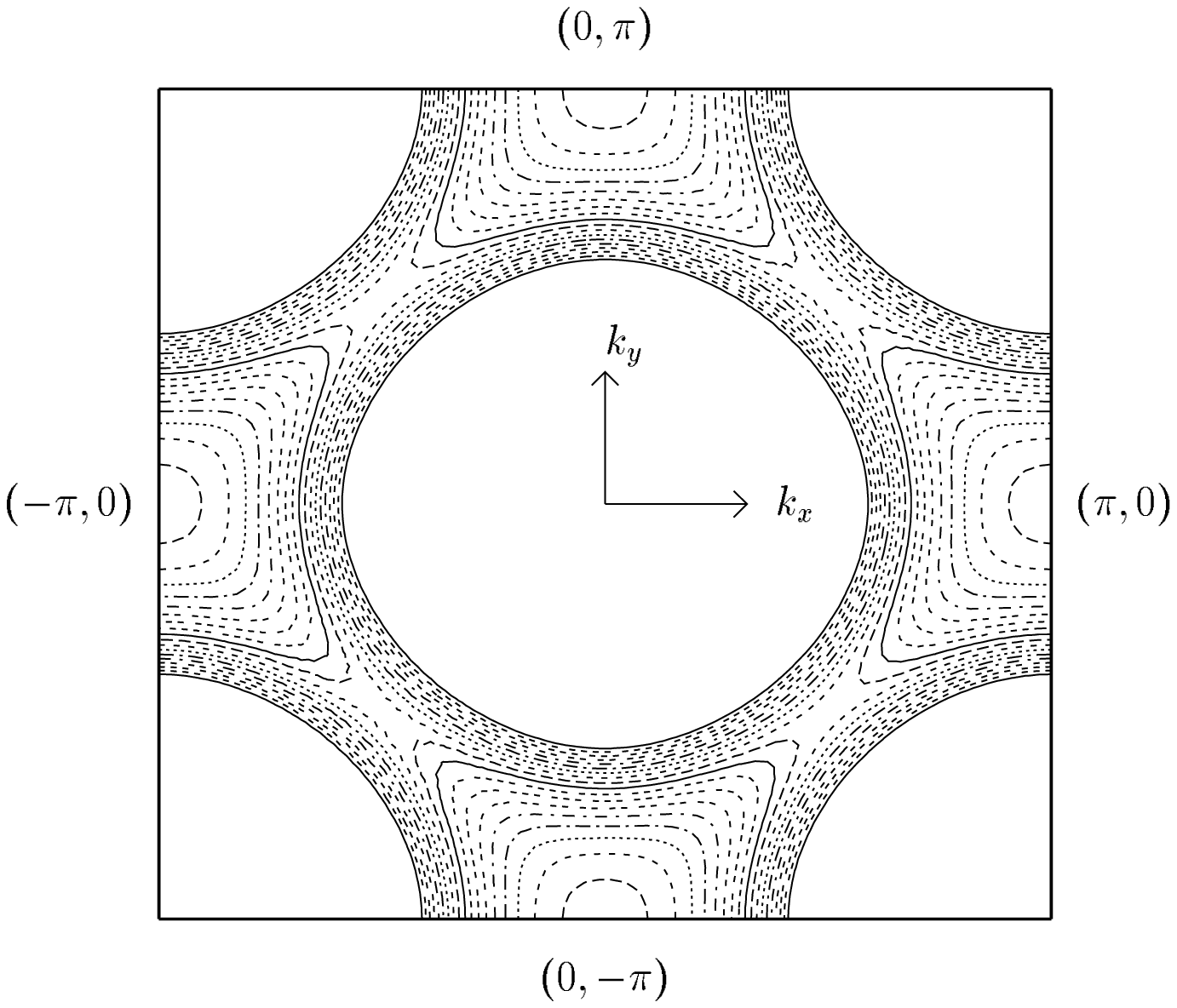,width=135mm,angle=0}
\vspace{-55mm}
\caption{Constant energy surfaces for the AF band energy
$E_{\bf k}^{\ominus}=\epsilon' _{\bf k} 
- (\Delta^2 + \epsilon_{\bf k}^2 )^{1/2} $,
showing the nearly (semi) circular contours
near the top of the lower energy band around the ${\bf k}$ points 
$(\pm \pi,0)$ and $(0,\pm \pi)$. Here $t'=0.05$ and $\Delta=5$.}
\end{figure}
\noindent
parametrized in terms of the angle $\theta$ shown in Fig. 3, 
so that $k_x=-(\pi- a\cos \theta)$ and $k_y=a \sin \theta$.
Thus, 
$\cos k_x \approx -1 + (a^2/2)\cos ^2 \theta$ and 
$\cos k_y \approx 1 - (a^2/2)\sin ^2 \theta$ up to O($a^2$),
so that $\epsilon_{\bf k}=-2t (a^2/2) \cos 2\theta \sim tx$
whereas $\epsilon'_{\bf k} = 4t' (1-a^2/2) \sim t'$.
Therefore, in the strong coupling limit 
the AF band energy $E_{\bf k}^{\ominus} \approx 
\epsilon'_{\bf k} - \epsilon_{\bf k}^2 /2\Delta - \Delta$,
which further simplifies to 
$\epsilon'_{\bf k} - \Delta$ in the limit $Jx^2 << t'$.
This yields a nearly quadratic energy dispersion
for $\bf k$ close to $(\pm \pi,0)$ or $(0,\pm \pi)$,
which accounts for the nearly circular constant-energy surfaces
and the flat density of states 
near the top of the lower band, as seen in Fig. 1.

For negative $t'$ and hole doping, the hole pockets in the lower band 
develop around the points $(\pm \pi/2,\pm\pi/2)$.
Translating the ${\bf k}$-space origin to $(\pi/2,\pi/2)$, 
and then rotating the coordinate system by $\pi/4$, 
the quasiparticle energy $E_{\bf k}^{\ominus}$ 
can be written in terms of the transformed coordinates
$\boldmath \kappa$ as 
\begin{equation}
-E_{\bf k}^{\ominus}=
2(J-|t'|)\; \kappa_x ^2 \;\; + \;\; 2|t'|\;  \kappa_y ^2
\end{equation}
to second order in $\kappa_x,\kappa_y$,
indicating that for $J>|t'|$ 
the constant-energy surfaces are ellipses centered at 
$(\pi/2,\pi/2)$.
Here the strong coupling limit has been taken, 
and the band energy shifted by $\Delta$.
Equating the total normalized area of the four elliptical hole pockets
to the doping concentration $x$, 
we obtain the semi-major and semi-minor axes of the Fermi-ellipse 
\begin{equation}
a = \left ( \frac{J-|t'|}{|t'|} \right )^{1/4} \sqrt{\pi x} 
\;\; , \;\;
b = \left ( \frac{|t'|}{J-|t'|} \right )^{1/4} \sqrt{\pi x} \; ,
\end{equation}
yielding the same $\sqrt{x}$ dependence 
as obtained earlier for the Fermi-circle radius for positive $t'$. 

\section{Stability of the AF state}
To test the stability of the AF state with respect to 
transverse perturbations $\Delta_\perp$ in the order parameter,
we examine the transverse response 
within first-order perturbation theory. 
For a transverse perturbation $\{\Delta_\perp ^i \}$,
which leads to terms $\Delta_\perp ^i 
a_{i\uparrow} ^\dagger  a_{i\downarrow} +
(\Delta_\perp ^i )^{\ast}
a_{i\downarrow} ^\dagger  a_{i\uparrow} $ in the Hamiltonian, 
the first-order corrections to the wavefunctions 
of the HF Hamiltonian $H(\Delta_z)$ are first obtained.
Next, the transverse mean-field potential 
$\tilde{\Delta}_\perp ^i = -U
\langle a_{i\downarrow} ^\dagger a_{i\uparrow} \rangle $
developed due to these corrections  
is obtained upto first order in $\Delta_\perp $;
this may be expressed as 
$\tilde{\Delta}_\perp ^i  = U \sum_j [A]_{ij} \Delta_\perp ^j $,
where the transverse response matrix $[A]$ 
connects the cause and effect.
If the largest eigenvalue $\lambda_\perp ^{\rm max}$ of this
matrix $[A]$ is larger (smaller) than $1/U$, then 
within a self-consistent approach, the AF state
characterized by the single order parameter $\Delta_z$ is 
unstable (stable) with respect to spin twisting.

It is shown in the Appendix that the  
transverse response matrix $[A]$ is nothing but the bare 
antiparallel-spin particle-hole propagator $[\chi^0]$ 
in the static limit ($\omega =0$).
In the AF state, where translational symmetry holds within 
the two-sublattice basis, Fourier transformation 
to wave vector (${\bf q}$) space reduces
$[\chi^0]$ to a $2\times 2$ matrix $[\chi^0({\bf q})]$
having two eigenvalue branches, which are further discussed below.

For the undoped $t-t'$-Hubbard model in the strong coupling limit, 
the AF insulating state involves 
a completely filled lower band and empty upper band, 
so that only {\em interband} particle-hole processes contribute
to $[\chi^0({\bf q})]$.
The relevant eigenvalue branch (having the larger eigenvalues)
is given by $\lambda_\perp ^{\rm inter}({\bf q}) = 
1/U - (t^2 / 4 \Delta^3)(1 - 2J'/J) q^2$ for small $q$.\cite{phase}
The corresponding eigenvector is $(1 \; -1)$, 
indicating that the transverse fluctuations 
$\Delta_\perp ^i = (-1)^i \Delta_\perp ^q e^{i{\bf q}.{\bf r}}$ 
for the mode $\bf q$ 
involve opposite sign on the two sublattices.
This implies that the mode $\bf q$ simply 
represents a spiral twisting of spins with an appropriate
wavelength. 
For $J'<J/2$, the eigenvalue is less than $1/U$ for all finite $q$, 
indicating the stability of the AF state, 
whereas for $J'>J/2$, the AF state is unstable towards a F-AF state
with ordering wavevector ${\bf Q}=(\pi,0)$,
possessing antiferromagnetic (ferromagnetic) 
ordering in the $x$ ($y$) direction.\cite{phase}
The $q=0$ mode corresponds to a rigid rotation of all spins,
and the marginal response for this mode $(U\lambda_\perp =1)$
merely reflects the equivalence of all 
spontaneous symmetry-breaking directions,
which follows from the Goldstone theorem
for systems possessing continuous symmetry. 
  
With doping the Fermi energy moves into a quasiparticle band,
and particle-hole excitations now include {\em intraband} processes 
involving states of the same band. 
By including these intraband processes in the 
transverse response matrix $[\chi^0]$, 
it was shown earlier for the doped Hubbard model that 
the eigenvalue {\em increases} beyond $1/U$ for finite $q$,  
indicating that the AF state  
is unstable towards the spiral twisting of spins 
for any finite doping.\cite{spiral3}
As the Hubbard model exhibits particle-hole symmetry,
this instability applies to both hole and electron doping.

In this section we show that in the $t-t'$-Hubbard model
this intraband contribution is strongly suppressed for positive
(negative) $t'$ and hole (electron) doping,
resulting in a stabilization of the AF state for a range of doping
concentration. This suppression is due to the modification
in the quasiparticle energy spectrum by $t'$,
which leads to a significant reduction in the 
phase space available for intraband processes.
However, for negative $t'$ and hole doping,
the AF state is found to remain unstable with respect to
transverse perturbations in the order parameter for any doping.

In order to obtain the transverse eigenvalue $\lambda_\perp$,
we evaluate, in the static limit ($\omega =0$),
the bare antiparallel-spin particle-hole propagator, 
$[\chi^0 ({\bf q},\omega)]=
i\int \frac{d\omega}{2\pi} \sum_{\bf k}'
[G^\uparrow ({\bf k}\omega')][G^\downarrow ({\bf k-q},\omega'-\omega)]$
in the AF state, separately focussing on the intraband and interband processes.
For the case of hole (electron) doping, 
intraband processes involve  
particle and hole states from the lower (upper) band, 
while interband processes involve particle and hole states from
different bands.

Now, the matrix $[\chi^0 ({\bf q})]$ has two eigenvalue branches
$[\lambda_\perp({\bf q})]^{1(2)}= 
[\chi^0 ({\bf q})]_{\rm AA} 
\pm [\chi^0 ({\bf q})]_{\rm AB} $,
corresponding to the two eigenvectors 
$( 1 \; \; 1 ) $ and $( 1 \; -1 ) $, respectively.
Separating the interband and intraband contributions 
as $[\chi^0 ({\bf q})] = 
[\chi^0 ({\bf q})]^{\rm inter} + 
[\chi^0 ({\bf q})]^{\rm intra}$,
the total transverse eigenvalues can be written as
$[\lambda_\perp({\bf q})]^{1(2)}= 
 [\lambda_\perp ^{\rm inter}({\bf q})]^{1(2)}  + 
 [\lambda_\perp ^{\rm intra}({\bf q})]^{1(2)}$.
We focus on the $( 1 \; \; -1 )$ branch which yields 
larger eigenvalues for small $q$, even when intraband processes 
are included.

In the following we show that for the $( 1 \; \; -1 )$ branch
the intraband contribution can be written as 
$\lambda_\perp ^{\rm intra}({\bf q})= 
(t^2/4\Delta^3)\alpha^{\rm intra} q^2$
for small $q$,
where $\alpha^{\rm intra}$ is a dimensionless positive coefficient.
The interband contribution can similarly be written as
$\lambda_\perp ^{\rm inter}({\bf q})= 
1/U - (t^2/4\Delta^3)\alpha^{\rm inter} q^2$.
The total transverse eigenvalue is therefore obtained as
\begin{eqnarray}
\lambda_\perp({\bf q}) &=& 
\lambda_\perp ^{\rm inter}({\bf q}) + 
\lambda_\perp ^{\rm intra}({\bf q}) \nonumber \\
&=& \frac{1}{U} - \frac{t^2}{4\Delta^3}
(\alpha^{\rm inter}- \alpha^{\rm intra}) q^2 \; .
\end{eqnarray}
For $q=0$ the transverse eigenvalue exactly equals $1/U$,
as required from the Goldstone theorem  
and the continuous spin-rotation symmetry.
Thus the question of stability of the AF state with respect to
transverse perturbations is reduced to a
comparison of the two coefficients $\alpha^{\rm intra}$ 
and $\alpha^{\rm inter}$.
If $\alpha^{\rm intra}$ exceeds $\alpha^{\rm inter}$, 
the transverse response 
$U \lambda_\perp({\bf q})  > 1$ for finite $q$, 
signalling a spiral instability.

\subsection{Intraband contribution}
In the two-sublattice basis (labelled by A,B),
the intraband contributions to the diagonal and off-diagonal  
elements are obtained as
\begin{eqnarray}
& & [\chi^0 ({\bf q},\omega)]_{\rm AA} ^{\rm intra}  \nonumber \\
&=& 
\sum_{\bf k} ' \left [
\frac
{
\left (a_{{\bf k}\uparrow}^\ominus \right )^2
\left (a_{{\bf k-q}\downarrow}^\ominus \right )^2
}
{
E_{\bf k-q}^\ominus - E_{\bf k}^\ominus + \omega -i\eta 
}
+
\frac
{
\left (a_{{\bf -k+q}\uparrow}^\ominus \right )^2
\left (a_{{\bf -k}\downarrow}^\ominus \right )^2
}
{
E_{\bf -k+q}^\ominus - E_{\bf -k}^\ominus - \omega -i\eta
}
\right ] 
\nonumber \\
& & [\chi^0 ({\bf q},\omega)]_{\rm BB} ^{\rm intra} \nonumber \\
&=&
\sum_{\bf k} ' \left [
\frac
{
\left (b_{{\bf k}\uparrow}^\ominus \right )^2
\left (b_{{\bf k-q}\downarrow}^\ominus \right )^2
}
{
E_{\bf k-q}^\ominus - E_{\bf k}^\ominus + \omega -i\eta 
}
+
\frac
{
\left (b_{{\bf -k+q}\uparrow}^\ominus \right )^2
\left (b_{{\bf -k}\downarrow}^\ominus \right )^2
}
{
E_{\bf -k+q}^\ominus - E_{\bf -k}^\ominus - \omega -i\eta
}
\right ]  \nonumber \\
& & [\chi^0 ({\bf q},\omega)]_{\rm AB}^{\rm intra} \;\;\; = \; \;\;
[\chi^0 ({\bf q},\omega)]_{\rm BA}^{\rm intra}  \nonumber \\
& = & \sum_{\bf k} ' \left [
\frac
{
a_{{\bf k}\uparrow}^\ominus b_{{\bf k}\uparrow}^\ominus
a_{{\bf k-q}\downarrow}^\ominus b_{{\bf k-q}\downarrow}^\ominus
}
{
E_{\bf k-q}^\ominus - E_{\bf k} ^\ominus + \omega - i \eta
}
+
\frac
{
a_{{\bf -k+q}\uparrow}^\ominus b_{{\bf -k+q}\uparrow}^\ominus 
a_{{\bf -k}\downarrow}^\ominus b_{{\bf -k}\downarrow}^\ominus
}
{
E_{\bf -k+q}^\ominus - E_{\bf -k} ^\ominus -\omega -i \eta
} \right ] , \nonumber \\
&  & 
\end{eqnarray}
where $\sum_{\bf k} ' $ indicates that
states ${\bf k, -k}$ are below the Fermi energy $E_{\rm F}$, while
states ${\bf k-q, -k+q}$ are above $E_{\rm F}$.

Simplification results from the property that the 
band energies $E_{\bf k}$ and the amplitudes 
$(a_{\bf k} \; b_{\bf k})$ are both unchanged under the 
transformation ${\bf k} \rightarrow {\bf -k}$.
The strong coupling limit $U >> t$ leads to further simplification
and in the static limit $(\omega=0)$ we obtain
for the diagonal 
$[\chi^0 ({\bf q})]_{\rm d} \equiv 
[\chi^0 ({\bf q})]_{\rm AA} =
[\chi^0 ({\bf q})]_{\rm BB} $
and off-diagonal  
$[\chi^0 ({\bf q})]_{\rm od} \equiv 
[\chi^0 ({\bf q})]_{\rm AB} =
[\chi^0 ({\bf q})]_{\rm BA} $
matrix elements 
\begin{eqnarray}
& &[\chi^0 ({\bf q})]_{\rm d}^{\rm intra} =
\sum_{\bf k} '
\frac
{
(\epsilon_{\bf k} ^2 + \epsilon_{\bf k-q} ^2 )/4\Delta^2
}
{
(\epsilon_{\bf k-q} ' - \epsilon_{\bf k} ' ) -
(\epsilon_{\bf k-q} ^2 - \epsilon_{\bf k} ^2 )/2\Delta
}
\nonumber \\
& &[\chi^0 ({\bf q})]_{\rm od}^{\rm intra} = 
\sum_{\bf k} '
\frac
{
2\epsilon_{\bf k}  \epsilon_{\bf k-q} /4\Delta^2
}
{
(\epsilon_{\bf k-q} ' - \epsilon_{\bf k} ' ) -
(\epsilon_{\bf k-q} ^2 - \epsilon_{\bf k} ^2 )/2\Delta
} . 
\end{eqnarray}

From  Eq. (9) the intraband eigenvalue
$\lambda_\perp ^{\rm intra}({\bf q}) =
[\chi^0 ({\bf q})]_{\rm AA}^{\rm intra} - 
[\chi^0 ({\bf q})]_{\rm AB}^{\rm intra} $
for the relevant $( 1 \; \; -1 ) $ branch is obtained as 
\begin{equation}
\lambda_\perp ^{\rm intra}({\bf q}) = \sum_{\bf k} '
\frac
{
(\epsilon_{\bf k} - \epsilon_{\bf k-q} )^2 /4\Delta^2
}
{
(\epsilon_{\bf k-q} ' - \epsilon_{\bf k} ' ) -
(\epsilon_{\bf k-q} ^2 - \epsilon_{\bf k} ^2 )/2\Delta
}\; .
\end{equation}
The relabelling transformation
$k_x \rightarrow -k_x$ and/or $k_y \rightarrow -k_y$ 
must leave
$\lambda_\perp ^{\rm intra}({\bf q})$ invariant,
and as both $\epsilon_{\bf k}$ and $\epsilon_{\bf k} ' $
are even functions of $k_x$ and $k_y$,
$\lambda_\perp ^{\rm intra}({\bf q})$ must also be an even function
of $q_x$ and $q_y$. Furthermore, since the $x$ and $y$ 
directions are equivalent, 
$\lambda_\perp ^{\rm intra}({\bf q})$ must vary 
as $q_x ^2 + q_y ^2$ to lowest order in $q$.
As $\lambda_{\rm intra}({\bf q})$ depends only on the magnitude 
$q$, we consider $q_y=0$ and $q_x = q$ for simplicity.
Figure 3 shows the hole Fermi circle of radius $a$ near the $(-\pi,0)$ point 
in the Brillouin zone. Consider another semi-circle whose 
center is shifted by the wave vector $q$, as shown.
For $\bf k$ states lying in the crescent region between these two semi-circles, 
we have $E_{\bf k}^\ominus <  E_{\rm F}$

\begin{figure}
\hspace*{-60mm}
\psfig{file=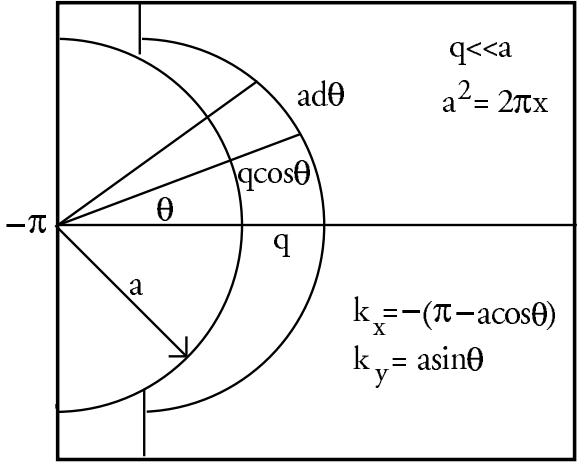,width=135mm,angle=0}
\vspace{-20mm}
\caption{
The ${\bf k}$ states which contribute to 
the intraband particle-hole processes near the $(-\pi,0)$ point 
lie in the crescent-shaped region.}
\end{figure}
\noindent
and $E_{\bf k-q}^\ominus > E_{\rm F} $,
and therefore it is this region of $\bf k$ space
which contributes to the intraband processes in Eq. (10).
As discussed earlier, for $q << a$, the Fermi circle radius,  
these $\bf k$ states can be parametrized as
$k_x=-(\pi- a\cos \theta)$ and $k_y=a \sin \theta$.
As $a^2 = 2\pi x$, for small doping concentration we have $a << 1$.

We now show that the two terms in the energy denominator in Eq. (10) 
are in the ratio $t' : Jx$, 
which plays a key role in determining the intraband contribution. 
Taylor expansion to first order yields
$\epsilon_{\bf k-q} - \epsilon_{\bf k}
=-{\bf q} . {\bf \nabla} \epsilon_{\bf k} 
=-2t q \sin k_x $ and
$\epsilon'_{\bf k-q} - \epsilon'_{\bf k}
=-{\bf q} . {\bf \nabla} \epsilon'_{\bf k} 
=-4t' q \sin k_x \cos k_y $ for small $\bf q$.
Therefore, we have
$\epsilon_{\bf k-q} - \epsilon_{\bf k} =
2tq \sin (a\cos \theta) \approx 2tq a\cos \theta$
for $a<<1$ and 
$\epsilon'_{\bf k-q} - \epsilon'_{\bf k} =
4t' q  \sin (a\cos \theta) \cos (a \sin \theta) \approx
4t' q  a\cos \theta $. 
As $\epsilon_{\bf k} \sim t x$,
it follows that the ratio of the two terms
in the energy denominator is $\sim t'/Jx$.
In the $t' << Jx$ limit, when the NNN hopping term may be 
dropped in comparison, it was shown that 
$\lambda_\perp ^{\rm intra}({\bf q})$ 
actually diverges as $x \rightarrow 0$,
so that the AF state 
is unstable for {\em any} doping concentration.\cite{spiral3} 

However, for $t' >> Jx$ 
the exchange-energy term may be dropped in Eq. (10).
Integrating over the crescent-shaped region shown in Fig. 3, 
and multiplying by 2 to account for the additional contributions
from the regions near $(0,\pm\pi)$, we obtain 
\begin{eqnarray}
\lambda_\perp ^{\rm intra}({\bf q}) &=& 
\frac{t^2}{4\Delta^3}
\frac{2\Delta}{t'}
\int_{-\frac{\pi}{2}} ^{\frac{\pi}{2}}
\frac{a \, d \theta \; q \cos \theta }{(2\pi)^2}
\;
\frac{\sin(a\cos\theta)}{\cos(a\sin\theta)}
\; q \; \nonumber \\
&\approx & 
\frac{t^2}{4\Delta^3}
\frac{\Delta}{t'}
\frac{a^2}{2\pi}
\int_{-\frac{\pi}{2}} ^{\frac{\pi}{2}}
\frac{d \theta}{\pi} \cos^2 \theta 
\; q^2 \nonumber \\
& = & 
\frac{t^2}{4\Delta^3}
\frac{\Delta x}{2t'} \; q^2  \; .
\end{eqnarray}
Comparison with Eq. (7) yields 
$\alpha^{\rm intra} = \Delta x / 2 t'$, 
which {\em vanishes} as $x \rightarrow 0$, 
in sharp contrast to the result for $t' << Jx$.
A similar result for $\alpha^{\rm intra}$
proportional to $x$ was obtained earlier.\cite{chub}

To evaluate $\lambda_{\perp}^{\rm intra}({\bf q})$
for negative $t'$ and hole doping,
we take $q_x=q_y$ for simplicity, 
so that ${\bf q} = (\sqrt{2}q,0)$ in the momentum coordinate
system rotated by $\pi/4$.
Substituting 
$E_{\bf k - q}^{\ominus} - E_{\bf k}^{\ominus}
\approx  4\sqrt{2}(J-|t'|)\kappa_x q $ from Eq. (5)
and 
$\epsilon_{\bf k-q} - \epsilon_{\bf k} \approx 4 t q$ in
Eq. (10), we obtain

\begin{equation}
\lambda_{\perp}^{\rm intra}({\bf q})
= \sum_{\boldmath \kappa} \frac {(16t^2/4\Delta^2) q^2}
{4\sqrt{2}(J-|t'|)\kappa_x q } \; .
\end{equation}
As the phase space available for the intraband process
is of order $aq$, and $\kappa_x$ is of order $b$,
we obtain 
\begin{equation}
\alpha^{\rm intra} \sim 
\frac{\Delta}{\sqrt{(J-|t'|)|t'|}} \; .
\end{equation}
In the strong coupling limit $\alpha^{\rm intra} >> 1$,
and is therefore much larger than $\alpha^{\rm inter} \sim 1$, 
so that the transverse response eigenvalue increases with $q$,
and the AF state is unstable for any doping concentration,
as also found in ref. [17].
This result is confirmed in the numerical study 
described in section IV, where it is seen that in the $t'-U$ region 
where the AF state exhibits longitudinal stability,
the maximum transverse response eigenvalue remains well above $1/U$
for all $U$.

\subsection{Interband contribution}
The interband eigenvalue $\lambda_{\perp}^{\rm inter}({\bf q})$ 
is next obtained for $t'>0$ 
by evaluating the interband contribution to $[\chi^0(\bf q)]$,
involving ${\bf k}$ states such that $E_{\bf k}^{\ominus} < E_{\rm F}$ 
and $E_{\bf k-q}^{\oplus} > E_{\rm F}$. 
The second condition holds for all ${\bf k}$ in the strong coupling limit, 
and the first condition is met by excluding in the ${\bf k}$ sum 
states lying within the four Fermi circles near $(\pm \pi,0)$ and
$(0,\pm \pi)$, as shown in Fig. 2.
Following the evaluation of the matrix elements of $[\chi^0(\bf q)]$
for the undoped $t-t'$-Hubbard model in the strong coupling 
limit,\cite{phase} 
and evaluating the contribution from within the Fermi circles (radius $a$)
up to order $a^4$,
we obtain  

\begin{eqnarray}
& & [\chi^0({\bf q})]_{AA}^{\rm inter} = 
[\chi^0({\bf q})]_{BB}^{\rm inter} = \frac{1}{U}  \nonumber \\
&-& \frac{t^2}{\Delta^3} \left[ \left \{ 1 - 
\frac{J'}{J}\left (1-\frac{\Delta}{t'}\frac{a^2}{2\pi}\right ) 
(1-\gamma' _{\bf q}) \right \} \right. \nonumber \\
&-&  \left \{ 
\left (\frac{a^2}{2\pi}-\frac{a^4}{8\pi}\right )
(\cos q_x -\cos q_y)^2
+\frac{a^4}{8\pi}(\sin^2 q_x + \sin^2 q_y) \right \} \nonumber \\
&+& \left \{ \frac{2J'}{J}
\left ( \frac{a^2}{2\pi}-\frac{a^4}{4\pi} \right )
(\cos q_x \cos q_y -1 )^2 \right \} \nonumber \\
&+& \left. \left \{ \frac{2J'}{J} 
\frac{a^4}{8\pi}(\cos^2 q_x \sin^2 q_y + \cos^2 q_y \sin^2 q_x)
\right \} \right ] \; , \nonumber \\ 
& & [\chi^0({\bf q})]_{AB}^{\rm inter} = 
[\chi^0({\bf q})]_{BA}^{\rm inter} = 
-\frac{t^2}{\Delta^3} \gamma_{\bf q} \; .
\end{eqnarray}
The off-diagonal elements are unchanged to order $a^4$,
the leading correction being of order $a^6$.
\ \\
\ \\
\ \\

Retaining terms only up to O$(a^2)$ and O($q^2$),
we obtain $\lambda_{\perp}^{\rm inter}({\bf q})$ 
corresponding to the $(1 \; \; -1)$ branch 
\begin{equation}
\lambda_{\perp}^{\rm inter}({\bf q})
= \frac{1}{U}
- \frac{t^2}{4\Delta^3} 
\left [ 1 - 
\frac{2J'}{J}\left (1-\frac{\Delta x}{t'}\right ) \right ] q^2 \; ,
\end{equation}
where the doping concentration $x=a^2/2\pi$. In the undoped case,
the $q^2$ term changes sign when $J' > J/2$, reflecting the
instability of the AF state towards the F-AF state with ${\bf Q}=(\pi,0)$.
Doping is seen to effectively reduce this frustrating effect of $t'$.

The phase boundary of the region of stability of the AF state,
obtained by equating the two coefficients $\alpha^{\rm inter}$
and $\alpha^{\rm intra}$ from Eqs. (11) and (15), 
is therefore given by 
\begin{equation}
1 - \frac{2J'}{J}\left (1-\frac{\Delta x}{t'}\right ) =
\frac{\Delta x}{2t'} \; .
\end{equation}
This is a cubic equation in $t'$ which may be written as 
\begin{equation}
f(t')\equiv \frac{4t'^{3} - 2t'}{4t'^{2} - 1} = \Delta x \; .
\end{equation}
The function $f(t')$ has two branches, as shown in Fig. 4,
yielding two solutions for any $\Delta x$. 
The first branch increases as $2t'$ for $t' << 1$,
and diverges as $t'\rightarrow 1/2$.
The second branch starts from $t'=1/\sqrt{2}$,
and asymptotically approaches $t'$ for $t' >> 1$.
This implies that for small $\Delta x$, 
the two solutions are $t'_{\rm min} \approx \Delta x/2$ 
and $t'_{\rm max} \approx 1/\sqrt{2}$, 
and the AF state is stable for $t'_{\rm min} < t' < t'_{\rm max}$.
Furthermore, $t'_{\rm min}$ is bounded by 1/2 from above, 
$t'_{\rm max}$ is bounded by $1/\sqrt{2}$ from below,
and $t'_{\rm max}$ asymptotically approaches $\Delta x$ for
large $\Delta x$.

To summarize, for fixed $x$,  the minimum $t'$ required
to stabilize the AF state increases in proportion
to the interaction  strength as $t'_{\rm min} \approx U x /4$,
and for fixed $t'$, the instability to a transverse perturbation develops 
when $ x > x_c $, 
where the critical doping concentration $x_c = 2 t'/\Delta \approx 4t'/U$.
Moreover, as $t'_{\rm max}$ increases away from $1/\sqrt{2}$ with increasing doping,
we find that hole doping interestingly suppresses the $t'$-induced frustration and stabilizes the AF state against the instability towards the F-AF state. 
These qualitative behaviours are indeed confirmed in the numerical study
described in the next section. For a quantitative comparison we take
a specific case $x=0.09$ (18 holes in a $14\times14$ lattice),
and $\Delta=mU/2 \approx (1-x)U/2 \approx 0.9$ for $U=20$,
so that $\Delta x \approx 0.8$. Graphical solution of Eq. (17)
by considering the intersections in Fig. 4 yields $t'_{\rm min} \approx 0.3$ and
$t'_{\rm max} \approx 1.1$, in excellent agreement with the
numerical result.

\begin{figure}
\vspace*{-70mm}
\hspace*{-38mm}
\psfig{file=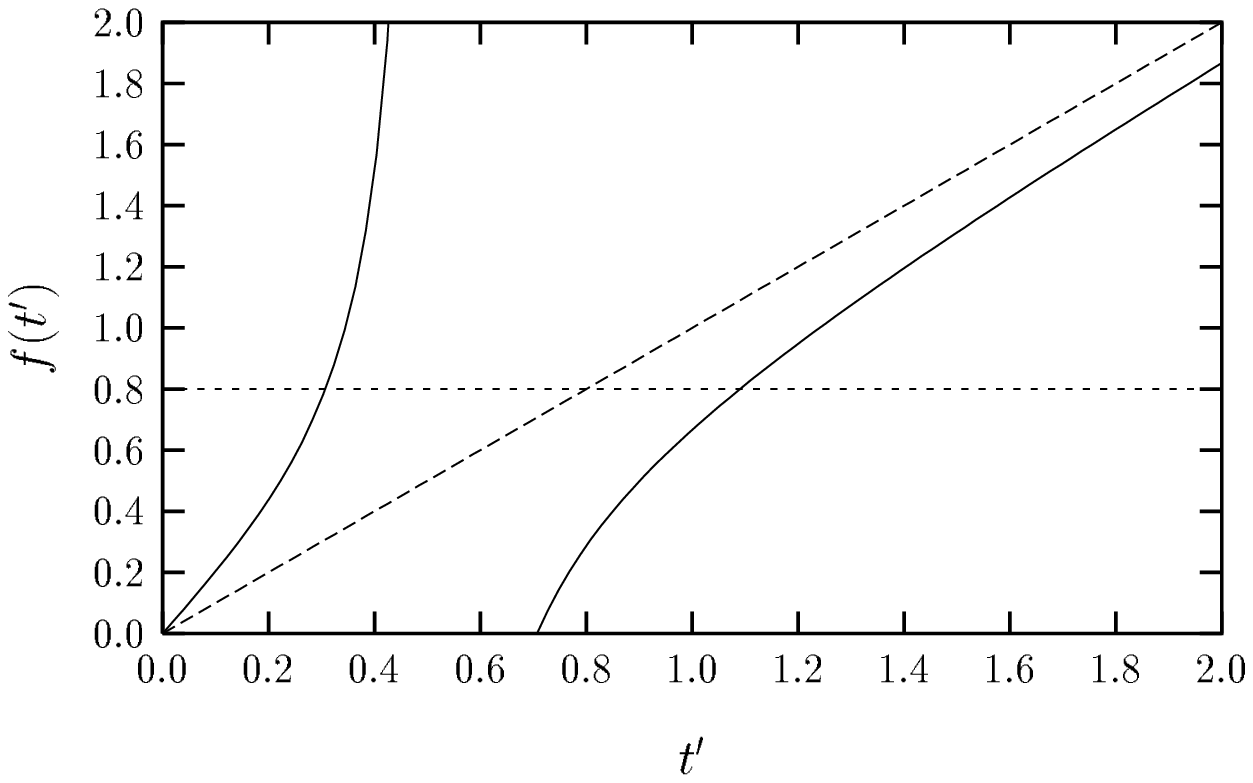,width=135mm,angle=0}
\vspace{-75mm}
\caption{
Plot of the function $f(t') = (4t'^3 -2t')/(4t'^2 -1)$ defined in Eq. (17)
showing the two branches.
Intersections with $\Delta x$ yield the two bounds  
$t'_{\rm min}$ and $t'_{\rm max}$, between which the AF state is stable.
For $\Delta x=0.8$, $t'_{\rm min} \approx 0.3$ and
$t'_{\rm max} \approx 1.1$.}
\end{figure}

\section{Phase Diagram}
A collinear mean-field state such as the AF state,
with spin ordering along some fixed (say, z) direction, 
can be unstable with respect to either 
longitudinal or transverse perturbations.
Longitudinal perturbations modify the local 
magnetization $\langle S_z ^i \rangle $ 
and therefore the local gap $\Delta_z ^i $,
whereas transverse perturbations induce spin twisting,
resulting in transverse terms $\Delta_\perp ^i $.
In this section we describe a numerical procedure for
studying the stability of the AF phase with respect to both these
perturbations, and discuss results for a $14\times 14$ lattice.
The objective is to obtain the $t' - U$ phase diagram   
for a fixed doping concentration, showing the different 
regions of stability/instability of the doped antiferromagnet. 

The usual numerical self-consistency procedure 
(with a spin-diagonal mean-field Hamiltonian 
corresponding to ordering in the z-direction)
is implicitly sensitive to instability with respect to longitudinal
perturbations due 
to the presence of small noise associated with the
numerics. Thus stability with respect to longitudinal fluctuations
is automatically indicated, at least for the finite-size lattice,
if a {\em homogeneous} AF state is obtained self-consistently.
Instability with respect to transverse perturbations
is studied by numerically evaluating the 
largest eigenvalue $\lambda_\perp ^{\rm max}$ 
of the transverse response matrix $[\chi^0]$.
Numerical evaluation of $[\chi^0]$, using the 
electronic eigenvalues and eigenfunctions 
of the self-consistent AF state, has been described earlier.\cite{st}
The largest eigenvalue $\lambda_\perp ^{\rm max}$ exceeding $1/U$
signals the transverse instability.

\begin{figure}
\vspace*{-60mm}
\hspace*{-38mm}
\psfig{file=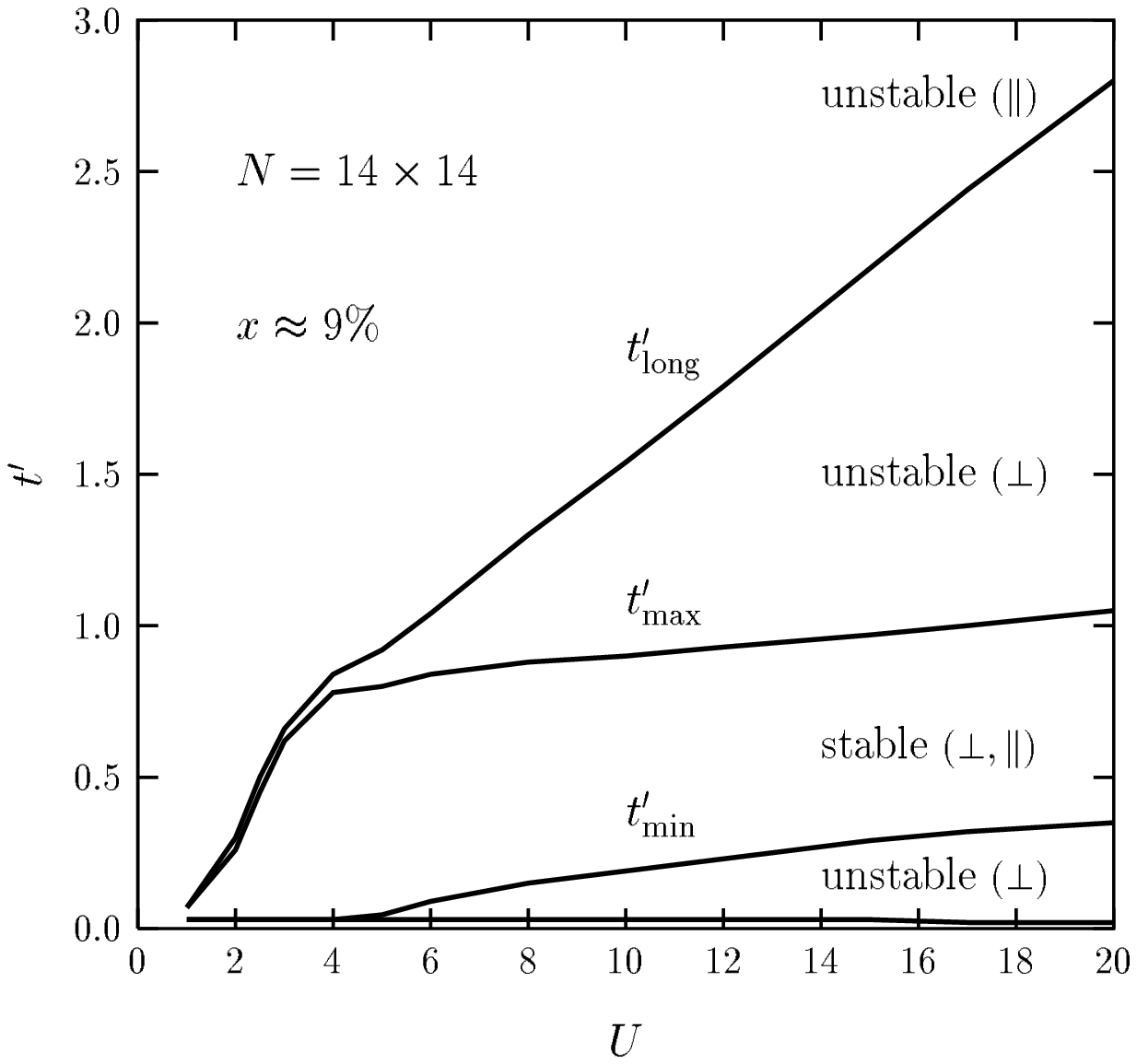,width=135mm,angle=0}
\vspace{-60mm}
\caption{
The  $t'-U$ phase diagram for positive $t'$ and hole doping 
of around 9\%, showing the region of stability 
between $t'_{\rm min}$ and $t'_{\rm max}$, where the AF state is 
stable with respect to both longitudinal ($\parallel$) and
transverse ($\perp$) perturbations.}
\end{figure}

\subsection{$t' > 0$}
Figure 5 shows the $t'-U$ phase diagram 
for a $14\times 14$ lattice with a fixed doping
concentration of around 9\% (nine holes per spin).
The minimum value $t'_{\rm min}$ above which the AF state
is stabilized is indeed found to increase nearly linearly
with $U$ in the strong coupling limit, and asymptotically approaches 0.5
with increasing $U$.
The stable region (with respect to both perturbations) 
lies in a band between $t'_{\rm min}$ and $t'_{\rm max}$.
The instability for $t' > t'_{\rm max}$ is 
towards the ${\bf Q}= (\pi,0)$ state, and $t'_{\rm max}$ increases 
away from $1/\sqrt{2}$ in the strong coupling limit.
These features are in agreement with the analytical results 
of the previous section.
With further increase in $t'$ the band gap continually decreases, 
causing an enhancement in the longitudinal response.
This eventually results in a longitudinal instability at 
$t' =  t'_{\rm long}$ where the gap vanishes.
As expected, $t'_{\rm long}$ 
scales linearly with $U$ in the strong coupling limit.
Longitudinal instability is also seen for $t' \lesssim 0.03$, 
essentially independent of $U$.

\subsection{$t' < 0$}
For negative $t'$ the AF-state energy eigenvalues in the lower band are 
four-fold degenerate, 
therefore the number of doped holes is taken in multiples of four.
The region of stability of the AF state 
with respect to longitudinal perturbations is shown in Fig. 6
for 8 doped holes per spin ($x\approx 8\%$).
For $t'$ values lying between the two curves
the homogeneous AF state is self-consistently obtained.
\begin{figure}
\vspace*{-70mm}
\hspace*{-38mm}
\psfig{file=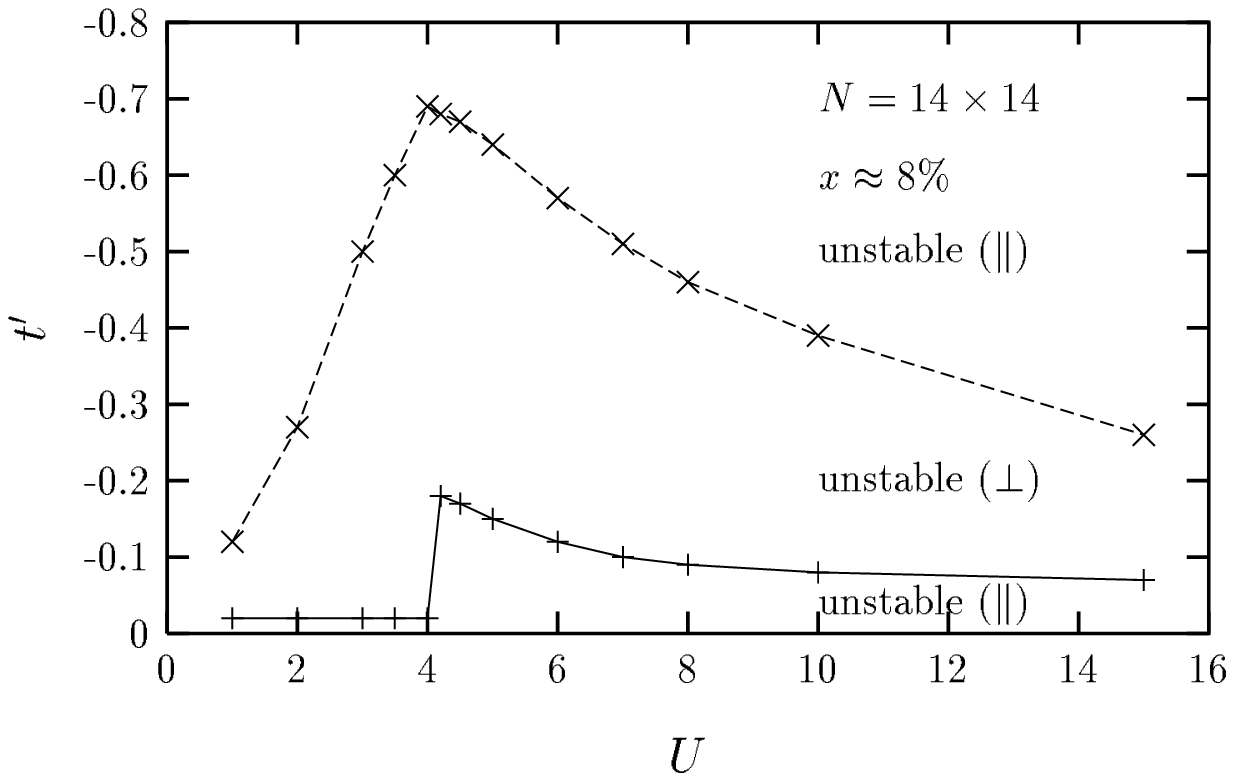,width=135mm,angle=0}
\vspace{-70mm}
\caption{
For negative $t'$ the phase diagram shows no region of stability for 
8 doped holes per spin ($x\approx 8\%$).
The AF state is stable with respect to longitudinal perturbations only
(in the region between the two curves).
The crossover at $U\approx 4$ coincides with the vanishing of the AF band gap.}
\end{figure}

\begin{figure}
\vspace*{-70mm}
\hspace*{-38mm}
\psfig{file=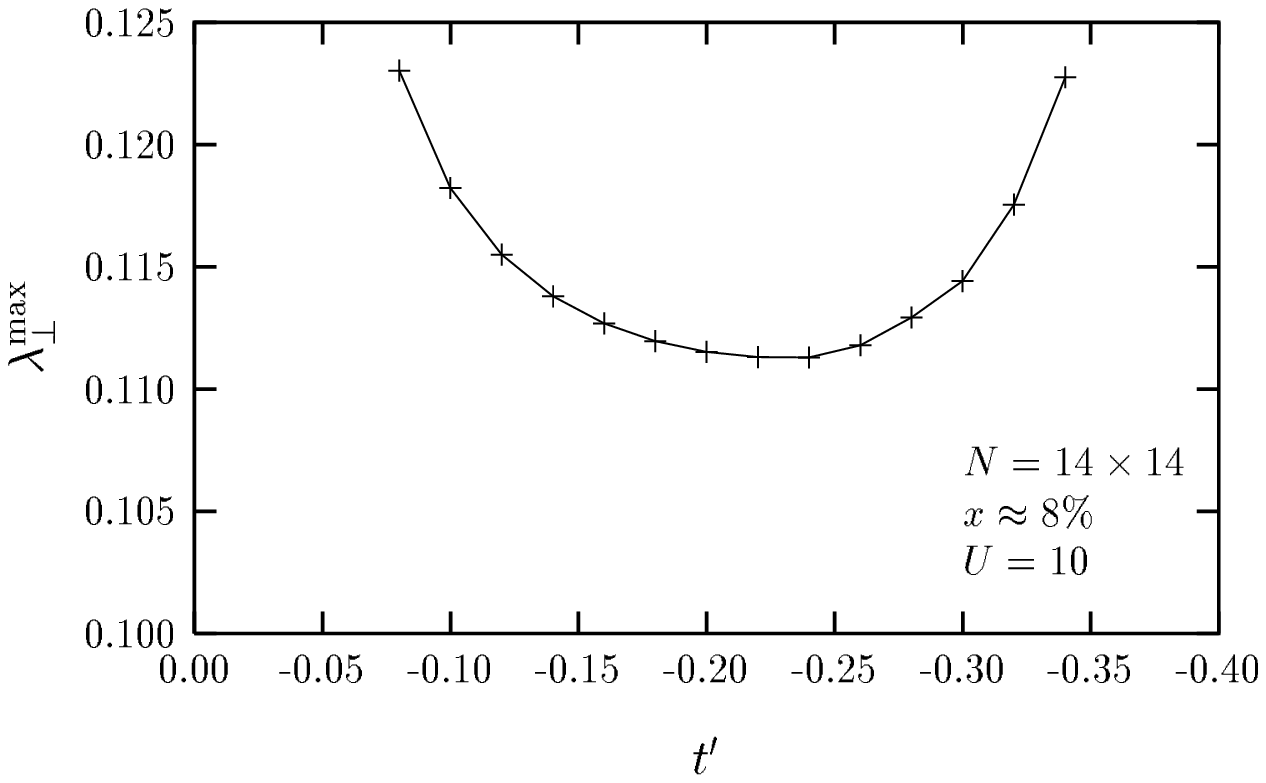,width=135mm,angle=0}
\vspace{-75mm}
\caption{
For negative $t'$, the largest eigenvalue $\lambda_\perp ^{\rm max}$ 
remains well above $1/U$, indicating no region of stability for the AF state
with respect to transverse perturbations.}
\end{figure}
\noindent
However, the response to transverse perturbations shows that
the AF state is actually unstable. This instability is 
reflected in the behaviour of the largest eigenvalue 
$\lambda_\perp ^{\rm max}$ of the $[\chi^0]$ matrix, 
shown in Fig. 7 for the specific case of $U=10$. 
It is seen that with increasing $|t'|$,
$\lambda_\perp ^{\rm max}$ approaches $1/U$ from above, but saturates
well above and starts increasing again. 
Similar behaviour is seen for intermediate $(U=5)$ and weak ($U=3$) coupling, 
and also for smaller doping concentration of 4 holes per spin ($x \approx 4\%$).
This leads to the conclusion that for negative $t'$, the hole-doped $t-t'$-Hubbard 
model does not possess a stable AF ground state.

\section{Transverse spin fluctuations}
The stable mean field AF state for the positive (negative)
$t'$ model and hole (electron) doping
yields a metallic AF state, characterized by 
intraband (gapless) particle-hole excitations, 
in addition to the interband excitations. 
In this section we study the magnon propagator,
given by $\chi^{-+}({\bf q},\omega)=
\chi^0 ({\bf q},\omega)/1-U \chi^0 ({\bf q},\omega)$
at the RPA level, 
which represents transverse spin fluctuations
about the broken-symmetry AF state.
This study should be especially relevant to
the electron-doped cuprate $\rm Nd_{2-x}Ce_x Cu O_4$.
We focus on the magnon damping,
which is associated with the 
imaginary part of $\chi^0 ({\bf q},\omega)$,
and arises from magnon decay into intraband particle-hole excitations.
Magnon damping is expected to play a crucial role
in determining the energy dependence of 
the local spin susceptibility 
$\sum_{\bf q} {\rm Im} \chi^{-+}({\bf q},\omega)$
in the doped antiferromagnet,
which is probed in neutron-scattering experiments.\cite{neutron}

\subsection{Evaluation of $[\chi^0 ({\bf q},\omega)]$ }
Here we examine the full ${\bf q},\omega$ dependence of  
$[\chi^0 ({\bf q},\omega)]^{\rm intra}$ in the AF state,
given in Eq. (8).
Considering first the imaginary part,  
we obtain for the AA matrix element 
\begin{eqnarray}
&&{\rm Im} [\chi^0 ({\bf q},\omega)]_{\rm AA} ^{\rm intra}  =
\sum_{\bf k} 
\theta(E_{\rm F} - E_{\bf k}^\ominus)\; 
\theta(E_{\bf k-q}^\ominus - E_{\rm F}) \nonumber \\
&&\pi \left [
\frac{\epsilon_{\bf k-q} ^2}{4\Delta ^2} \;
\delta ( E_{\bf k-q}^\ominus - E_{\bf k}^\ominus + \omega )
+
\frac{\epsilon_{\bf k} ^2}{4\Delta ^2} \;
\delta ( E_{\bf k-q}^\ominus - E_{\bf k}^\ominus - \omega )
\right ]. \nonumber \\
& & 
\end{eqnarray}
Here terms of the type $\epsilon_{\bf k} ^2/4\Delta ^2$
are the AF coherence factors, 
and the $\theta$ functions ensure that 
states $\bf k$ lie below the Fermi energy
while ${\bf k-q}$ lie above it.
The ${\bf k}$ states which contribute in the above 
expression lie in the crescent-shaped region of Fig. 3.

In the $t' >> Jx$ limit considered earlier, 
and for the ${\bf k}$ states of interest,
we have
$E_{\bf k-q}^\ominus - E_{\bf k}^\ominus \approx
\epsilon'_{\bf k-q} - \epsilon'_{\bf k}
\approx 4t'q a \cos \theta $ 
and $\epsilon_{\bf k} \approx -2t (a^2/2) \cos 2\theta$
for small $q_x = q$ and vanishing $q_y$.
Substituting in Eq. (18), and taking $\omega > 0$, we obtain 
\begin{eqnarray}
& &{\rm Im} [\chi^0 ({\bf q},\omega)]_{\rm AA} ^{\rm intra}  =
2 \pi \int_0 ^{\pi/2} 
\frac{a d\theta \; q \cos \theta}{(2\pi)^2}
\nonumber \\
& & \frac{4t^2}{4\Delta^2}
\left ( \frac{a^2}{2} \right ) ^2 \cos ^2 2\theta \;
\delta (4t'q a \cos \theta - \omega)
\nonumber \\
&=& \frac{t^2}{4\Delta^3} 
\frac{\Delta a^4}{8\pi t'}
\left [
\frac{\cos\theta}{\sin\theta} \cos ^2 2 \theta 
\right ]_{4t'q a \cos \theta = \omega}
\end{eqnarray}
for $\omega < 4t'q a$. 
The delta function never clicks for $\omega >  4t' q \, a $
and the imaginary part vanishes.
Thus a new energy scale $\omega_x \equiv 4t' q \, a$ 
characterizes the imaginary part of 
${\rm Im} [\chi^0 ({\bf q},\omega)]$, and therefore the 
spin fluctuation spectrum. 

\begin{figure}
\vspace*{-70mm}
\hspace*{-38mm}
\psfig{file=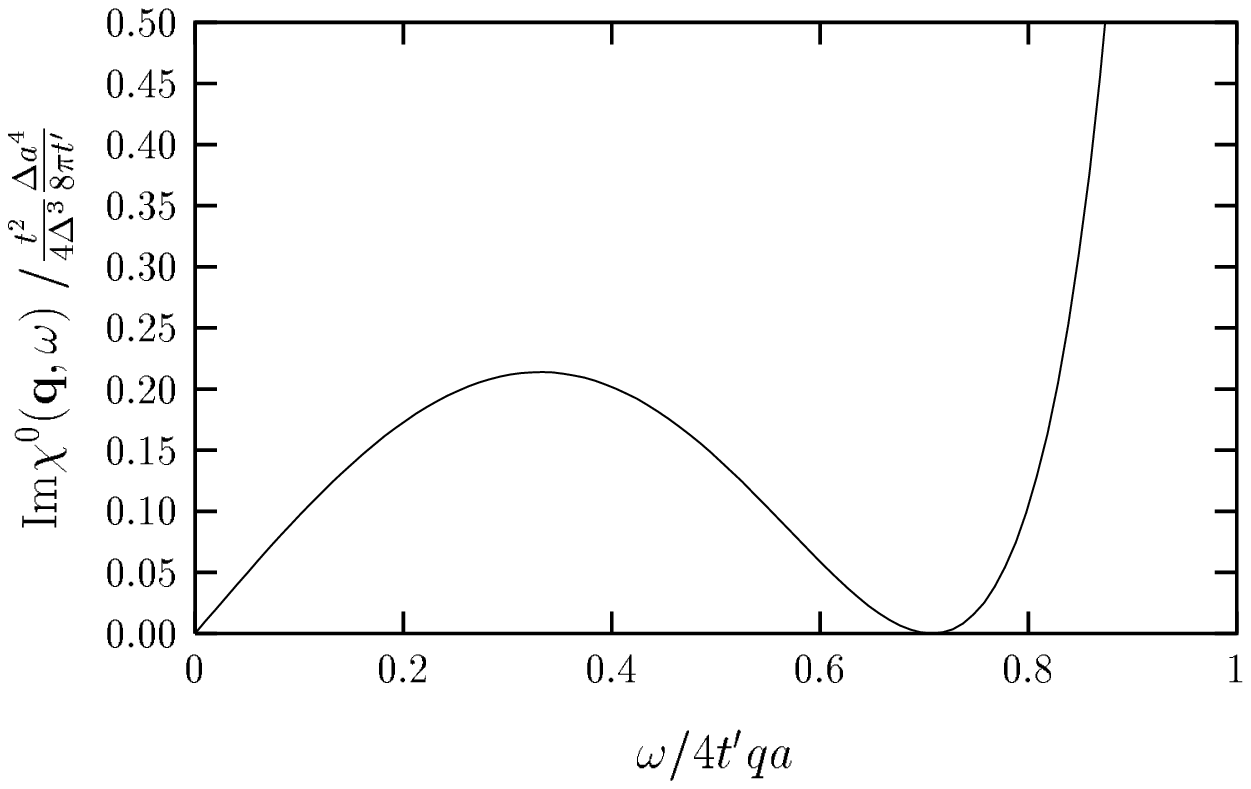,width=135mm,angle=0}
\vspace{-70mm}
\caption{
Plot of the imaginary part ${\rm Im}\chi^0 ({\bf q},\omega)$ vs. 
$\omega/4t' q a$, showing the linear dependence on $\omega$
for $\omega/4t' q a << 1$.}
\end{figure}
\noindent

The $\omega$ dependence is particularly simple in the low-energy limit
$\omega << \omega_x$. In this limit $\theta \approx \pi/2$,
so that $\sin \theta \approx 1$ and $\cos ^2 2 \theta \approx 1$,
and therefore we obtain
\begin{equation}
{\rm Im} [\chi^0 ({\bf q},\omega)]_{\rm AA} ^{\rm intra}  =
\frac{t^2}{4\Delta^3} 
\frac{\Delta a^4}{8\pi t'}
\left (\frac{\omega}{\omega_x} \right )
\end{equation}
which scales like $\Delta x^2 /t'$ and increases linearly with $\omega$.
On the other hand, as $\omega \rightarrow \omega_x $ from below, 
$\cos \theta \rightarrow 1$ in Eq. (19), and  therefore
${\rm Im} [\chi^0 ({\bf q},\omega)]$  diverges like 
$1/\sqrt{1- \omega^2/\omega_x ^2 }$. 
At intermediate energies, when $\omega \approx \omega_x /\sqrt{2}$,
we have $\theta \approx \pi/4$ so that the imaginary part 
nearly vanishes due to the $\cos^2 2\theta$ term, 
which originates from the AF coherence factor.
The full $\omega$ dependence is shown in Fig. 8.

We now consider the other matrix elements of 
${\rm Im} [\chi^0 ({\bf q},\omega)]$.
For the off-diagonal term we obtain
(again for $\omega > 0$)
\begin{eqnarray}
&{\rm Im}& [\chi^0 ({\bf q},\omega)]_{\rm AB} ^{\rm intra}  =
\sum_{\bf k} 
\theta(E_{\rm F} - E_{\bf k}^\ominus)\; 
\theta(E_{\bf k-q}^\ominus - E_{\rm F}) \nonumber \\
& & \pi \left [
\frac{\epsilon_{\bf k} \epsilon_{\bf k-q }}{4\Delta ^2} \;
\delta ( E_{\bf k-q}^\ominus - E_{\bf k}^\ominus - \omega )
\right ] \; .
\end{eqnarray}
The relevant AF coherence factor in this case is
$\epsilon_{\bf k} \epsilon_{\bf k-q }/4\Delta^2$ 
instead of $\epsilon_{\bf k} ^2/4\Delta^2 $. 
Similarly the B-sublattice term
${\rm Im} [\chi^0 ({\bf q},\omega)]_{\rm BB}$
involves the factor $\epsilon_{\bf k-q } ^2 /4\Delta^2$.
Now $\epsilon_{\bf k-q } \approx \epsilon_{\bf k } $
in the limit of small $q$, 
therefore to leading order in $q$ all the matrix elements yield
the same contribution obtained above.
Including this intra-band contribution 
to the imaginary part of $[\chi^0 ({\bf q},\omega)]$, 
we obtain

\begin{equation}
[\chi^0 ({\bf q},\omega)]= \frac{1}{U} {\bf 1} - \frac{t^2}{\Delta ^3}
\left [ \begin{array}{lr} 
1 -i\, \Gamma^0 + \omega & \gamma_{\bf q} - i\,  \Gamma^0 \nonumber \\
\gamma_{\bf q} - i\,  \Gamma^0  & 1 -i\, \Gamma^0 - \omega
\end{array}
\right ]\; .
\end{equation}
Here the energies $\omega$ and $\Gamma^0$ are measured in units of
the magnon energy scale $2J$, and the imaginary term $\Gamma^0$ is given
by 
$(t^2/\Delta^3) (\Gamma^0/2J) = 
{\rm Im} [\chi^0 ({\bf q},\omega)]^{\rm intra}$,
given in Eqs. (19) and (20). 
In the above equation we have retained only the 
intra-band contributions to the imaginary terms and
have neglected those for the real terms, which will be small
in the limit of small doping concentration. 
\subsection{Magnon Propagator}
Substituting in the RPA expression 
$[\chi^{-+} ({\bf q},\omega)] = 
[\chi^0 ({\bf q},\omega)]/{\bf 1} -U[\chi^0 ({\bf q},\omega)]$
for the magnon propagator, we obtain
\begin{eqnarray}
[\chi^{-+} ({\bf q},\omega)] 
&=& 
\left [ \begin{array}{lr} 
(1 -i\, \Gamma^0 -\omega) & -(\gamma_{\bf q} - i\,  \Gamma^0) \nonumber \\
-(\gamma_{\bf q} - i\,  \Gamma^0)  & (1 -i\, \Gamma^0 +\omega)
\end{array}
\right ] \nonumber \\
& \times & 
\frac{1}{\omega_{\bf q} ^2 - \omega^2 - 2i\Gamma^0 (1-\gamma_{\bf q} ) } 
\end{eqnarray}
where $\omega_{\bf q} = \sqrt{1-\gamma_{\bf q} ^2} $ 
is the bare magnon energy (in units of $2J$) 
for the two-dimensional antiferromagnet.
In the small momentum limit $\gamma_{\bf q} \approx 1$, 
so that in the energy denominator above the imaginary term  
$2i\Gamma^0 (1-\gamma_{\bf q}) \approx 
i\Gamma^0 (1-\gamma_{\bf q}^2 ) = i\Gamma^0 \omega_{\bf q}^2$,
and the magnon propagator expression simplifies to
\begin{eqnarray}
[\chi^{-+} ({\bf q},\omega)] &=&
\left [ \begin{array}{lr} 
(1 -i\, \Gamma^0 -\omega) & -(\gamma_{\bf q} - i\,  \Gamma^0) \nonumber \\
-(\gamma_{\bf q} - i\,  \Gamma^0)  & (1 -i\, \Gamma^0 +\omega)
\end{array} \right ] 
\frac{1}{(1-i\Gamma ^0 /2) }  \nonumber \\
& \times &
\left ( \frac{1}{\omega - \omega_{\bf q} + i \Gamma } -
        \frac{1}{\omega + \omega_{\bf q} - i \Gamma } \right )
\frac{1}{2\omega_{\bf q} }
\end{eqnarray}
where O$(\Gamma^0)^{2}$ terms have been dropped.
The magnon damping term is finally obtained as  
$\Gamma = \Gamma^0 \omega_{\bf q} /2 $.
In the low-energy limit $\omega << \omega_x$, 
where $\Gamma^ 0$ is given by Eq. (20), the magnon damping 
becomes proportional to energy
\begin{equation}
\frac{\Gamma}{2J} \sim \frac {\omega}{J'/x^{3/2}} \; ,
\end{equation}
indicating that the characteristic energy scale for magnon damping
goes as $J' / x^{3/2}$, where $J'=4t'^{2} /U$ is the exchange energy
corresponding to the NNN hopping $t'$. 

As discussed earlier, 
for $\omega > \omega_x \equiv 4t'aq$, the imaginary term $\Gamma_0$ vanishes, 
and therefore with an infinitesimally small damping term 
($\Gamma \rightarrow 0$)
the magnon propagator in Eq. (24) yields a delta function for 
the spectral function ${\rm Tr}\,{\rm Im}\chi^{-+}({\bf q},\omega)$ 
at $\omega=\omega_{\bf q}$.
If $J > 4t'a $, then $\omega_{\bf q} > \omega_x$ for all the
small $q$ modes, 
and in this case the local transverse spin susceptibility
$\sum_{\bf q}{\rm Tr}\,{\rm Im}\chi^{-+}({\bf q},\omega)
\sim \int qdq \, \delta(\omega-\omega_{\bf q}) /\omega_{\bf q} = \; {\rm constant}$
for $\omega > 4t'a$.
However, in the low-energy limit ($\omega << \omega_x$),
evaluation of $\sum_{\bf q}{\rm Tr}\,
{\rm Im}\chi^{-+}({\bf q},\omega)$ from Eq. (24)
shows that the local spin susceptibility vanishes as $\omega \rightarrow 0$.
This result is qualitatively similar to the pseudogap behaviour
seen in the spin excitation spectrum of doped cuprates in
neutron-scattering experiments.\cite{neutron}

\section{Conclusions}
By significantly modifying the AF band energies near the 
top (bottom) of the lower (upper) band, 
a positive (negative) $t'$ term strongly suppreses the
intraband contribution to the transverse response eigenvalue,
resulting in a stabilization of the AF state for finite hole (electron) doping.
We have derived analytical expressions for the 
interband and intraband contributions to the transverse response eigenvalue 
in the strong coupling and weak doping limit,
yielding a cubic equation in $t'$ for the phase boundary, 
solutions of which yield the region of stability of the AF state.
We find that doping supresses the $t'$-induced frustration due to the 
competing interaction $J'$ and interestingly stabilizes the AF state 
against the instability towards the F-AF phase with ${\bf Q}= (\pi,0)$. 
These analytical results are in quantitative agreement with
a numerical stability analysis on finite lattices, 
which also yields the full phase diagram in the $t'-U$ space,
indicating regions of stability of the AF state with respect to both
transverse and longitudinal perturbations. 

For negative (positive) $t'$, however, the AF state is found to remain
unstable for any hole (electron) doping.
For a fixed sign of $t'$,
the location and shape of the Fermi surface,
the DOS near $E_{\rm F}$,
and therefore the intraband contribution,
are all very different for electron and hole doping,
as the two AF bands are highly asymmetric.
This accounts for the strong dependence of the stabilization 
on the type of doping, 
and qualitatively explains why AF ordering in the electron-doped cuprate 
$\rm Nd_{2-x}Ce_x Cu O_4$ is much more robust 
than in the hole-doped $\rm La_{2-x}Sr_x Cu O_4$.

Finally, the study of transverse spin fluctuations in the
metallic state of the doped AF yields an interesting
low-energy result $\Gamma \sim \omega$ for magnon damping,
arising from magnon decay into particle-hole excitations. 
This is of relevance 
in the context of phenomenological theories 
such as the nearly AF Fermi-liquid theory,
put forward to explain the Knight shift experiments and 
describe the non-Korringa temperature dependence 
of the nuclear spin-lattice relaxation rate in doped cuprates. 
As the AF-state energy spectrum is what essentially determines the 
imaginary part of the particle-hole propagator, 
charge fluctuations will also exhibit a similar $\omega/\omega_x$ dependence. 
A detailed study of spin and charge fluctuations in the metallic AF state,
and their consequences on the AF long-range order,
the local spin susceptibility 
$\sum_{\bf q} {\rm Im} \chi^{-+}({\bf q},\omega)$,
and the electronic spectrum is currently in progress.\cite{met_af}

\section{Appendix}
We consider a mean-field AF state with spin polarizations
in the z direction, so that the mean-field Hamiltonian
is spin diagonal. Let the eigenvalues and eigenfunctions
in this state be denoted by 
$\{ {\cal E}_{l\sigma} , |l \sigma \rangle \}$.
The state $|l\sigma \rangle $ has amplitudes $\phi_{l\sigma} ^i$
on site $i$ for spin $\sigma$ and zero amplitude for the 
opposite spin. 
Then the first-order correction to the
state $|l\downarrow \rangle $ 
due to a small transverse perturbation $\Delta_\perp$
is obtained as
\begin{equation}
|\delta l\uparrow \rangle  = 
\sum_{m\uparrow}
\frac
{|m\uparrow \rangle \langle m \uparrow 
|\Delta_\perp|
l\downarrow \rangle } 
{{\cal E}_{l\downarrow}-{\cal E}_{m\uparrow} } \; .
\end{equation}
The correction $|\delta l\uparrow \rangle $
yields spin-$\uparrow$ amplitudes generated by the
perturbation to the state $|l\downarrow \rangle $, which has only
spin-$\downarrow$ amplitudes.
A similar equation yields the correction 
$|\delta l\downarrow \rangle $ for the $|l\uparrow \rangle$ state.

Now, the transverse mean-field potential 
$\tilde{\Delta}_\perp ^i = -U 
\langle a_{i\downarrow} ^\dagger a_{i\uparrow} \rangle$
generated due to the transverse perturbation
can be derived from the exact eigenstates $|L\rangle$, and we obtain 
\begin{eqnarray}
\tilde{\Delta}_\perp  & = &  
 -U \sum_L ' \langle L|
\left [ \begin{array}{lr} 0 & 0 \\ 1 & 0 
\end{array} \right ] |L \rangle 
\nonumber \\
    & = &
-U \sum_l ' 
\langle l\downarrow | \delta l \uparrow  \rangle 
+ \langle \delta l\downarrow | l \uparrow  \rangle 
\end{eqnarray}
to first order in the perturbation. 
Here the $'$ indicates that the sum is over all occupied states.
Substituting for the amplitudes, we obtain
\begin{equation}
\tilde{\Delta}_\perp ^i  = 
-U \sum_j \sum_{lm} '
\left (
\frac
{
\phi_{l\uparrow}^i \phi_{m\downarrow}^i
\phi_{m\downarrow}^j \phi_{l\uparrow}^j
}
{
{\cal E}_{l\uparrow}-{\cal E}_{m\downarrow} 
}
+
\frac
{
\phi_{l\downarrow}^i \phi_{m\uparrow}^i
\phi_{m\uparrow}^j \phi_{l\downarrow}^j
}
{
{\cal E}_{l\downarrow}-{\cal E}_{m\uparrow} 
}
\right )
\Delta_\perp ^j
\end{equation}

As the two terms above 
are antisymmetric under exchange of labels $l$ and $m$, 
all terms with states $m$ below
the Fermi energy cancel pairwise, and we obtain
\begin{eqnarray}
\tilde{\Delta}_\perp ^i &=& 
U \sum_j \sum_{{\cal E}_{l} < {\cal E}_{\rm F}}
^{{\cal E}_{m} > {\cal E}_{\rm F}}
\left (
\frac
{
\phi_{l\uparrow}^i \phi_{m\downarrow}^i
\phi_{m\downarrow}^j \phi_{l\uparrow}^j
}
{
{\cal E}_{m\downarrow}-{\cal E}_{l\uparrow} 
}
+
\frac
{
\phi_{l\downarrow}^i \phi_{m\uparrow}^i
\phi_{m\uparrow}^j \phi_{l\downarrow}^j
}
{
{\cal E}_{m\uparrow}-{\cal E}_{l\downarrow} 
}
\right ) \Delta_\perp ^j \nonumber \\
&=& 
U \sum_j [\chi^0]_{ij} \Delta_\perp ^j \; .
\end{eqnarray}
Hence the transverse response matrix $[A] = [\chi^0]$, 
the bare antiparallel-spin particle-hole 
propagator in the static limit ($\omega =0$).

\ \\ 
\ \\
\ \\
\ \\

\end{multicols}
\end{document}